\documentstyle[12pt]{article}

\def\beq{\begin{equation}}
\def\eeq{\end{equation}}
\def\bea{\begin{eqnarray}}
\def\eea{\end{eqnarray}}
\def\bq{\begin{quote}}
\def\eq{\end{quote}}

\makeatletter
\def\vereq#1#2{\lower3pt\vbox{\baselineskip1.5pt \lineskip1.5pt
\ialign{$\m@th#1\hfill##\hfil$\crcr#2\crcr\sim\crcr}}}
\makeatother

\begin{document}

\begin{titlepage}
\begin{center}
\today     \hfill    SLAC-PUB-7777 \\
~{} \hfill CERN-TH/98-90  \\
~{} \hfill hep-th/9804068\\
 
\vskip .1in

{\large \bf Exact results for non-holomorphic masses in \\
softly broken supersymmetric gauge theories}

\vskip 0.1in

Nima Arkani-Hamed$^a$ and Riccardo Rattazzi$^b$

\vskip .05in

{\em $^a$ Stanford Linear Accelerator Center\\
Stanford University\\
Stanford, California 94309, USA}
\vskip 0.3truecm
{\em $^b$ Theory Division, CERN\\
CH-1211, Gen\`eve 23, Switzerland} 
\end{center}

\begin{abstract}
We consider strongly coupled supersymmetric gauge theories
softly broken by the addition of gaugino masses $m_\lambda$ 
and (non-holomorphic) scalar masses $m^2$, taken to be small relative to 
the dynamical scale $\Lambda$. 
For theories with a weakly coupled dual description in the infrared, we 
compute exactly
the leading soft masses for the ``magnetic" degrees of freedom,
with uncalculable corrections suppressed by powers of $(m_{\lambda}/\Lambda),
(m/\Lambda)$.
The exact relations hold between the infrared fixed point ``magnetic" 
soft masses and the ultraviolet fixed point ``electric'' soft masses,
and correspond to a duality mapping for soft terms. 
We briefly discuss implications of these results for 
the vacuum structure of these theories.
\end{abstract}

\end{titlepage}

Recent years have seen enormous progress in our understanding of strongly
coupled supersymmetric gauge theories \cite{Seib,seiwit}.
In particular, a large class of models
have ``dual" descriptions which are weakly coupled in terms of
``magnetic" degrees of freedom in the deep infrared.
It is natural to attempt to extrapolate these supersymmetric
results to non-supersymmetric theories by adding
soft masses for the superpartners in order to decouple them 
\cite{hsu,peskin,dhoker,alvarez,hsin,jimsteve,yest}.
As a modest first step towards the decoupling limit,
one can study the response of the theory
to soft masses $m$ much smaller than the dynamical scale
of the theory $\Lambda$.
 This is also of considerable interest
to models where some of the Standard Model fields arise as composites
of elementary ``preons". If
the preon soft masses are known, what are the soft masses of the composite
states? The difficulty in addressing these simple questions is that scalar
soft terms are given by manifestly non-holomorphic terms in the bare Lagrangian
\begin{equation}
{\cal L}_{soft} \supset \int d^4 \theta \, \theta^2 \bar{\theta}^2 m^2
\phi^{\dagger} e^{V} \phi
\end{equation}
and so the usually powerful constraint of holomorphy can not be used to
analyse this type of soft breaking in theories with $N=1$
supersymmetry. Nevertheless, in this letter we will show that
the leading contribution to the soft masses for
the ``magnetic" fields can be computed exactly in terms of the
soft masses for the original fields, with uncalculable
corrections suppressed by powers of $(m/\Lambda)$.
These results are possible due to an interpretation of scalar
soft masses as auxilliary components of the vector field of an
 anomalous background $U(1)$ gauge symmetry \cite{us}.
In ref. \cite{us}, this symmetry was exploited in perturbation theory,
allowing high-loop supersymmetry breaking results to be obtained
from lower loop {\it supersymmetric} computations. Here we
show that the same symmetry can be useful for computing soft masses
in strongly coupled theories.

For simplicity, we will consider (asymptotically free) SUSY gauge theories 
with a simple gauge group,
softly broken by a gaugino mass $m_\lambda$ and universal scalar masses $m^2$.
The extension of our methods to include several group factors as well 
as arbitrary scalar masses
will be obvious. Begin by considering the exactly supersymmetric limit.
The bare lagrangian with ultraviolet cutoff $\mu_{UV}$ is
\begin{equation}
\int d^2\theta S(\mu_{UV}) W^2+{\rm h.c}+
\int d^4\theta F(S(\mu_{UV})+ S^\dagger(\mu_{UV}))Q^\dagger Q
\label{susy}
\end{equation}
where $``Q^{\dagger} Q"$ stands for $Q^{\dagger} e^V Q$. 
The $\mu_{UV}$ dependence of 
$S,F$ is dictated by the Wilsonian 
renormalization group, which requires that the low energy physics stay fixed
as the ultraviolet cutoff $\mu_{UV}$ is varied. It is well-known that 
$S$ only changes at 1-loop \cite{SV}, whereas $F$ runs at all orders in 
perturbation theory: 
\begin{eqnarray}
\frac{d S }{d  \ln \mu_{UV}}  = -\frac{b}{8 \pi^2} \\
\frac{d \ln F}{d \ln \mu_{UV}} = \gamma(S)
\end{eqnarray}
where $\gamma$ is the supersymmetric anomalous dimension.

Now consider the reparametrization $Q \to {\sqrt {Z}}Q$.
By the rescaling anomaly \cite{kon,mehit2} the new lagrangian is
\begin{equation}
\int d^2\theta (S(\mu_{UV})+{T\over 8\pi^2}\ln Z) W^2+{\rm h.c}+
\int d^4\theta Z F(S(\mu_{UV}) + S^{\dagger}(\mu_{UV}))Q^\dagger Q.
\label{two}
\end{equation}
Where $T$ is the total Dynkin index of the matter fields $Q$. 
After relabelling $S(\mu_{UV})+(T/8\pi^2)\ln Z\to S(\mu_{UV})$,
we can start over again with the following bare lagrangian
\begin{equation}
\int d^2\theta S(\mu_{UV}) W^2+{\rm h.c}+
\int d^4\theta Z F(S(\mu_{UV}) + S^\dagger(\mu_{UV})
-{T\over 4 \pi^2}\ln Z)Q^\dagger Q.
\label{three}
\end{equation}
and treat $S$ and $Z$ as independent parameters.
The theory defined by eqn.(\ref{three}) 
is invariant under the transformation
\begin{equation}
Z\to Z\chi\chi^\dagger, \quad Q\to Q/\chi, \quad S(\mu_{UV}) \to 
S(\mu_{UV}) + (T/4\pi^2)\ln \chi.
\label{inv}
\end{equation}
Notice that the ``physical" coupling ${\rm Re}S-(T/4\pi^2)\ln Z$
is invariant. 

We can now consider the situation where $S(\mu_{UV})$ and $Z$ are 
respectively promoted to chiral and real vector superfields.
For us this just means that these quantities have non vanishing
$\theta^2$ and $\theta^2\bar \theta^2$ components, corresponding
to soft gaugino and squark masses. Notice that, when $S$ and $Z$ are promoted
to superfields, the bare lagrangian of eqn.(\ref{three}) still defines
a cut-off independent low energy theory also including insertions of
soft terms. This follows from  simple power counting. Apart from a
cosmological constant term, no new divergences are generated.
 Indeed, these would
have to involve covariant derivatives acting on $S$ and $Z$, but there
is no local counterterm of this type also involving
the physical fields.
 
Now that $S$ and $Z$ are superfields the 
above invariance becomes an abelian background $U(1)_A$ gauge symmetry.
Physical quantities have to be $U(1)_A$ and RG invariant.
The parameter space of the theory is described by the only $U(1)_A$
and RG invariant object that can be formed with $S$ and $Z$
\begin{equation}
I \equiv \Lambda_h^\dagger Z^{2T/b}\Lambda_h
\label{four}
\end{equation}
where $\Lambda_h=\mu_{UV} e^{-8\pi^2S/b}$ is the holomorphic 
dynamical scale.
The $\theta^0$ component
of $I$ gives  the ``physical'' strong scale 
$[I]_{\theta=\bar{\theta}=0} = \Lambda^2$ \cite{mehit}. 
As we show immediately below, the $\theta^2$ and $\theta^2\bar \theta^2$
components are related to the UV fixed point limits of the
gaugino mass $m_\lambda$ and the squark mass $m_Q^2$ respectively:
\bea
&&[\ln I]_{\theta^2} \equiv {16 \pi^2 \over b} m_g = \lim_{\mu_{UV}\to \infty}{16 \pi^2 \over b}
\left ({m_\lambda\over
g^2}\right )\nonumber \\
&&[\ln I]_{\theta^2\bar\theta^2} =  \frac{2T}{b} [\ln Z]_{\theta^2\bar\theta^2}
\equiv - \frac{2T}{b} m^2=-
\frac{2T}{b} \lim_{\mu_{UV}\to\infty} m_Q^2 
\label{limits}
\eea
As a first step in making these identifications, we show that in the deep 
UV $\mu_{UV} \to \infty$, the $\theta^2$ and $\theta^2 \bar{\theta}^2$ 
components of $F$ vanish and hence make no contribution to soft terms.
By dimensional analysis and invariance under the anomalous symmetry 
the wave function has the form $F=F(\mu_{UV}^2/I)$ (note that 
$-b/8 \pi^2$ ln $\mu_{UV}^2/I = S + S^\dagger - T/4 \pi^2$ ln $Z$ is 
just the argument of $F$ in eqn.(\ref{three})). Therefore we have
\bea
[\ln F]_{\theta^2}= - \frac{1}{2} 
\frac{d \ln F}{d \ln \mu_{UV}} [\ln I]_{\theta^2} 
= -\frac{8 \pi^2}{b} \gamma(\mu_{UV}) m_g,
\eea
\bea
[\ln F]_{\theta^2\bar\theta^2}&&=
-{1\over 2}{d \ln F\over d \ln\mu_{UV}}[\ln I]_{\theta^2\bar\theta^2}
+{1\over 4}{d^2 \ln F\over d \ln^2\mu_{UV}}|[\ln I]_{\theta^2}|^2 
\nonumber \\
&&={T\over b}\gamma(\mu_{UV})\, m^2+\left ({8\pi^2\over b}\right)^2\,
\dot \gamma(\mu_{UV})\,m_g^2
\label{Ftheta}
\eea
where $\dot\gamma = d \gamma/d$ln$\mu_{UV}$.
As $\mu_{UV} \to \infty$, the theory becomes free, $\gamma, 
\dot\gamma\to 0$, 
and $[\ln F]_{\theta^2},[\ln F]_{\theta^2\bar\theta^2}$ both vanish.

We now establish the first of eqns.(\ref{limits}).
Defining the anomalous $U(1)_A$ invariant quantity $R$ via \cite{SV}
\beq 
R - \frac{A}{8 \pi^2} \mbox{ln} R =S+S^\dagger-(T/4\pi^2)\ln ZF
\eeq
(where $A$ is the Dynkin index of the adjoint representation)
the physical gauge coupling
in any given scheme has the form \cite{us}
\begin{equation}
R_{\rm phys}=R+\sum_{n=0}^\infty c_nR^{-n}
\label{real}
\end{equation}
where the  $c_n$ are scheme dependent constants (the coefficient of 
the leading term
is fixed by the equality of Wilsonian and physical coupling at tree level).
At any scale $\mu_{UV}$, we have that $m_\lambda/g^2=[\ln R_{\rm phys}
]_{\theta^2}$. 
However, as $\mu_{UV} \to \infty$, 
$R^{-1}\to 0$ , so that only the first term in eqn. (\ref{real}) matters,
and the first of eqns.(\ref{limits}) follows trivially. The same result
was discussed in refs. \cite{hisano,us}. Consider now the running squark mass
given by the matter kinetic term in eqn.(\ref{three})
\begin{equation}
m_Q^2(\mu_{UV}) = -[\mbox{ln} Z]_{\theta^2 \bar{\theta}^2} - [\mbox{ln} \
F(\mu_{UV})]_{\theta^2 
\bar{\theta}^2}.
\label{mmu}
\end{equation}
Again, as $\mu_{UV} \to \infty$, the second 
term in eqn.(\ref{mmu}) vanishes and we recover the second of 
eqns.(\ref{limits}).

Having established the physical interpretation of the various components
of the $U(1)_A$ and RG invariant 
superfield $I$, we discuss the computation of ``magnetic" soft masses.
This will be possible since the anomalous $U(1)_A$ symmetry of eqn.(\ref{inv}) 
provides a powerful constraint on the way in which $S,Z$ 
(and hence the soft masses) enter into the theory. As an example, consider
$SU(N)$ SUSY QCD with $(N+1)$ flavors 
$Q_i,\bar{Q}_{\bar{i}}$, for the moment in the supersymmetric limit. 
In the deep infrared and at the 
origin in moduli space, this theory has a weakly coupled description in 
terms of the composite ``mesons" $M_{i \bar{i}} = Q_i \bar{Q}_{\bar{i}}$ 
and ``baryons" $B^{i} = (Q^N)^i,\bar{B}^i = (\bar{Q}^N)^{\bar{i}}$ \cite{Seib}.
We expect that, as long as the soft masses are much smaller than the strong
scale $\Lambda$, the mesons and the baryons still give a good
description of the low energy theory. In particular, we expect
the Kahler potential for these fields to be smooth 
everywhere on moduli space. 
Therefore we can expand it in a power series in $M,B,\bar B$ around the origin.
By using invariance under the flavor symmetries,  under
eqn.(\ref{inv}), and under the RG, the Kahler potential
must depend on $S,Z$ as
\begin{equation}
K = c_M \frac{M^{\dagger} Z^2 M}{I} + c_B \frac{B^{\dagger} Z^N B}{I^{N-1}} + 
c_{\bar{B}} \frac{\bar{B}^{\dagger} Z^N \bar{B}}{I^{N-1}} + \cdots
\label{kahler}
\end{equation}
The effective Kahler potential $K$ is also associated with a
coarse-graining scale $\mu_{IR}< \Lambda$, and the wave function 
coefficients $c_{M,B,\bar B}$ (which depend on $\mu_{IR}$) play a role 
similar to $F$ in the 
UV theory. At any $\mu_{IR}$, the soft terms for the composites are given 
by e.g.
\begin{equation}
m^2_M(\mu_{IR}) = -[\mbox{ln} \frac{Z^2}{I}]_{\theta^2 \bar{\theta}^2} - 
[\mbox{ln} c_M(\mu_{IR})]_{\theta^2 \bar{\theta}^2}.
\label{comp}
\end{equation} 
By invariance under the 
ultraviolet RG and the anomalous $U(1)$, the wave functions have the form
$c_{M,B,\bar B}\equiv c_{M,B,\bar B}(\mu_{IR}^2/I)$. As for the
UV wave function $F$, the dependence of the $c$'s on the soft terms
is determined by the RG
\begin{equation}
[\ln c_{M,B,\bar B}]_{\theta^2\bar\theta^2}=
{T\over b}\gamma_{M,B,\bar B}(\mu_{IR})\, m^2+\left ({8\pi^2\over b}\right)^2
\,\dot \gamma_{M,B,\bar B}\,(\mu_{IR})m_g^2
\label{ctheta}
\end{equation}
where, similarly as before, $\gamma=d\ln c/d\ln\mu_{IR}$ and $\dot 
\gamma=d^2\ln c/d\ln^2 \mu_{IR}$.
%
Now, the effective theory of mesons and baryons is free in the IR.
In fact it involves one marginal 
Yukawa interaction $W\supset \bar B M B$ which goes to zero 
for $\mu_{IR}\to 0$.
More precisely as ${\mu_{IR}\to 0}$, $c_{C,B,\bar B} \to \infty$,
so that the effective coupling 
$\lambda_{eff}(\mu_{IR})^2 \sim 1/c_M c_B c_{\bar B} \to 0$ and 
the anomalous dimensions $\gamma_{M,B,\bar{B}} \to 0$. We conclude 
that at $\mu_{IR}=0$ the $c$'s do not affect the soft terms in eqn.
(\ref{comp}).
We emphasize that this argument is completely analogous to the one given 
above for the irrelevance of $F$ to the soft terms in the deep UV. 
Eqn. (\ref{ctheta}) determines a relation between soft terms and RG 
which is somewhat similar to the
one discussed in ref. \cite{gianricc}. In that case the role
of the invariant $I$ was played by the messenger threshold
superfield $XX^\dagger$.

By the above discussion, the IR fixed point value of the 
composites are determined by the $\theta^2 \bar{\theta}^2$ components of 
$Z,I$ which are in turn related to the 
UV fixed point value of the squark masses as in eqn.(\ref{limits}). We 
therefore find a purely algebraic relationship between the
composite soft masses in the deep IR and the squark masses in the deep UV. 
Using the anomalous $U(1)$ symmetry, we can seemingly control the exact soft
masses for the composites, at least at the origin of moduli space! This is
however true only in the limit in which the soft masses
$m_g$ and $m$ are much smaller than the strong scale $\Lambda$. Indeed 
there are $U(1)_A$ invariant terms in $K$ which 
can involve the $U(1)_A$ ``field strength" 
$W^{\alpha}_{A} = 
\bar{D}^2 D^{\alpha} \ln Z$ which is non-vanishing when
ln$Z$ has a non-vanishing $\theta^2 \bar{\theta}^2$ component.
One such term is 
\begin{equation}
\int d^4 \theta \, \left(\frac{D_\alpha W^\alpha_{A}}{I} \right) 
{M}^{\dagger} {Z^2\over I}{M}
\end{equation}
Since
$W^\alpha_{A}$ has positive mass dimension, however,  
this and all other such operators make contributions to the composite soft 
masses which are 
suppressed by powers of $(m/\Lambda)$.
It is these uncontrollable operators which prevent us from taking the 
decoupling
limit $(m/\Lambda) \to \infty$, 
however, their effects are power suppressed for $(m/\Lambda) \ll 1$.

We have now all the ingredients to determine the mapping of soft terms
between the microscopic and macroscopic theories, up to corrections suppressed
by powers of $(m^2/\Lambda^2)$: 
\begin{eqnarray} 
m_M^2(\mu_{IR}=0) &=& -[\mbox{ln} \frac{Z^2}{I}]_{\theta^2 \bar{\theta}^2} =
{2N-4\over 2 N-1}\, m_Q^2(\mu_{UV}=\infty) \\ 
m_{B,\bar{B}}^2(\mu_{IR}=0) 
&=&-[\mbox{ln} \frac{Z^N}{I^{N-1}}]_{\theta^2 \bar{\theta}^2}= 
{2-N\over 2N-1}\, m_Q^2(\mu_{UV}=\infty).
\label{sun}
\end{eqnarray}
These masses satisfy the relation $m_M^2+m_B^2+m_{\bar B}^2=0$. This
sum rule can be inferred from the low energy theory due to the RG ``focusing'' 
effect of the Yukawa interaction $\bar B M B$, and could have been 
established without using
the anomalous symmetry\cite{hsin}. The symmetry is however crucial to fix the
value of each mass. Notice that for $N>2$ and for positive squark masses 
the baryons are tachyonic.
The implications of this result for  the symmetry properties of the
vacuum will be discussed below. Notice also that for the special case of 
$SU(2)$ the baryons and the mesons coincide and have vanishing soft mass.
While this result holds in the deep IR, by eqs.(\ref{comp}-\ref{ctheta}) we 
can establish
how this limit is approached. This is a pure Yukawa theory
for which both $\gamma_M$ and $\dot\gamma_M$
are negative in the perturbative domain. Therefore we conclude that $m_M^2$ 
is positive at finite $\mu_{IR}$ and approaches zero as $\mu_{IR} \to 0$.

The result in eqn. (\ref{sun}) has a nice interpretation in terms of the
anomalous $U(1)$ charges of the canonically normalized fields 
$\hat M=M/\Lambda_h$ and $\hat B=B/\Lambda_h^{N-1}$, ($\Lambda_h$
 has charge $2T/b$).
In terms of the canonical fields eqn. (\ref{kahler}) reads
\begin{equation}
K = c_M \hat M^{\dagger} Z^{q_{\hat M}} \hat M + c_B \hat B^{\dagger} 
Z^{q_{\hat B}} \hat B + 
c_{\bar{B}} \hat {\bar{B}}^{\dagger}Z^{q_{\hat B}} \hat{\bar{B}} + \cdots .
\label{canonical}
\end{equation}
The charges of the composites $q_{\hat M}$ and $q_{\hat B}$
coincide with the corresponding IR/UV mass ratios of eqn. (\ref{sun}).
Notice that, without the contribution to the soft terms from the powers of
$I$ in eqn. (\ref{kahler}),
the soft masses of the composites would just be determined in the ``naive''
way, by adding the masses of the constituents.

We stress that the existence of a relationship between deep UV and IR quantities is
just a consequence of RG invariance. For instance, in QCD one may ask for the
expression of the pion mass in terms of the fundamental parameters.
It will have the form $m_\pi^2=c \,\hat m_q \Lambda_{QCD}$, where $c$ 
is a constant
and $\hat m_q$ is an RG invariant combination of the running quark mass
$m_q(\mu)$ and gauge coupling $g^2(\mu)$. In practice, as for $m_g$ in eqn.(\ref{limits}),
one can define $\hat m_q$ just by using the 1-loop RG in the deep UV
 $\hat m_q=\lim_{\mu\to\infty} g^p(\mu) m_q(\mu)$, where $p$ is determined
by the 1-loop $\beta$ function and mass anomalous dimension.
The striking feature of our case, with respect to QCD, is that
we can calculate the analogue of the coefficient $c$.

Another example is given by $Sp(k)$ gauge theory with $2k+4=2N_F$ 
chiral multiplets $L_i$ 
in the fundamental representation ($SU(2)$ with 3 flavors
is just the special case $k=1$). The low energy description involves
the antisymmetric meson field $V_{ij}=L_i\epsilon L_j$ with a superpotential
$W_{\rm conf}=$Pf$V/\Lambda_h^{2k +1}$ 
\cite{pouliot}. By adding  soft terms the
resulting  mass for the meson is (from now on it is understood that the LHS 
and RHS
soft masses are the IR and UV fixed point values respectively)
\begin{equation}
m_V^2= {2k -1 \over 2k +1} \, m^2_L
\label{spmass}
\end{equation}
Notice that ${\rm Pf}\, V\sim V^{k+2}$ is an irrelevant operator in the low 
energy theory for $k>1$.
In this case $\gamma_V(\mu_{IR}^2/I)$ goes to zero with a power law when 
$\mu_{IR}\to 0$ and the ``running'' mass $m^2_V(\mu_{IR})$ approaches 
eqn.(\ref{spmass}) equally fast.

Finally consider $SU(N)$ gauge theory with $N+1<N_F<3N/2$ for which the
low-energy description is in terms of a dual ``magnetic'' theory with
gauge group $SU(N_F-N)$. The magnetic theory contains an elementary
meson $M_{i \bar{j}}$ and $N_F$ flavors of dual quarks $q^i, \bar 
q^{\bar i}$
in the fundamental representation of $SU(N_F-N)$. The $U(1)_A$ charge of the
canonically normalized 
meson $\hat M_{i \bar{j}}= M_{i \bar{j}}/\Lambda_h$ is just
$q_{\hat{M}}= q(Q_i\bar Q_{\bar j}/\Lambda_h)=2(3N-2N_F)/(3N-N_F)$. The charge
of the dual quarks simply follows from the invariance of the tree level
magnetic superpotential $W_{\rm magn}=\bar q^{\bar i} M_{\bar i j} q^j/
\Lambda_h$ 
(or by matching the baryons in the 
two theories $b=q^{N_F-N}=Q^N/\Lambda_h^{2N-N_F}=B/\Lambda_h^{2N-N_F}$). 
We thus obtaing the soft masses of the magnetic theory
\begin{equation}
m_M^2= 2{3N-2N_F\over 3N-N_F}\, m^2\quad\quad 
m_{q,\bar{q}}^2=-{3N-2N_F\over 3N-N_F}\,
m^2.
\label{sundual}
\end{equation}
Again the ``baryons'' $q, \bar q$ are tachyonic for all theories in 
the free magnetic phase $N_F<3N/2$. 
For $3N/2<N_F<3N$ the theory is in 
an interacting non-Abelian Coulomb phase. Here we cannot apply our method
in an obvious way since there are no points where the theory is free.
It is interesting that all the magnetic soft masses vanish at the boundary between the free
magnetic and conformal windows, $N_F = 3N/2$.

Notice that the normalization of the magnetic quarks $q,\bar q$ is arbitrary,
and that in ref. \cite{Seib} a scale $\mu$ was introduced to give
proper dimension to the superpotential: 
$W_{\rm magn} = M_{i \bar j} q^i \bar q^{\bar j}/\mu$. Correspondingly the 
holomorphic
scales in the electric and magnetic theory are related by \cite{Seib}
$\Lambda_h^b\tilde \Lambda_h^{\tilde b}=\mu^{N_F}$ (where the tilded quantities
refer to the magnetic theory). In our derivation we have 
fixed $\mu = \Lambda_h$, but our results do not depend on that choice.
Indeed one could have argued as follows. With the normalization of 
ref. \cite{Seib}
the dual quarks have charge $-1$ under the anomalous $U(1)_A$ 
($\mu$ is neutral
and $M_{i \bar{j}}=Q_i\bar Q_{\bar j}$).
As we did for the electric theory in eqn. (\ref{three}),
we can define a dual wave function $\tilde Z$ multiplying the kinetic term
of the dual quarks $q,\bar q$ as well as an invariant scale $\tilde I=
\tilde \Lambda_h^\dagger \tilde Z^{2T/\tilde b}\tilde \Lambda_h$. 
However, since the electric and magnetic theory describe the same physics,
it must be  $\tilde I=I$. Therefore we deduce the
following ``duality'' relation
\begin{equation}
\tilde Z^b\,= \, Z^{\tilde b}
\label{dualz}
\end{equation}
which holds up
to a gauge transformation which does not affect the  
mapping of soft terms. Eqn.(\ref{dualz}) correctly gives
the magnetic quark masses in eqn.(\ref{sundual}).
By eqn.(\ref{dualz}) the opposite sign of electric and magnetic squark masses
is a reflection of duality between IR and UV free theories.
Finally,
by considering  $[\ln \tilde I]_\theta^2=
[\ln I]_\theta^2$, a similar duality is obtained for gaugino
masses. More precisely one gets (notice again the flip in sign)
\begin{equation}
\lim_{\mu_{IR}\to 0} \,\left ({m_\lambda\over \tilde b g^2}\right)_{\rm magn}=
\lim_{\mu_{UV}\to \infty} \,\left ({m_\lambda\over b g^2}\right)_{\rm el}.
\end{equation}

Our results for the soft masses of composite and ``magnetic" fields 
have obvious implications for composite model-building in theories
where supersymmetry breaking is communicated to the ``preon" fields 
at a scale higher than the dynamical scale $\Lambda$ (for instance 
by supergravity mediation). Clearly one must check that e.g. 
none of the composite squarks obtain negative soft masses. 

Next we consider the vacuum structure of these theories, 
begining with the $SU(N)$ theories
with $N+1 \leq N_F \leq 3/2 N$ flavors and $N>2$. In the supersymmetric limit, these theories
have a moduli space of vacua, and we are interested in how the soft breaking effects lift this 
vacuum degeneracy. 
In all these theories, for positive 
squark masses in the deep ultraviolet, 
the ``meson" fields get positive 
soft masses while the ``baryonic" fields get negative soft 
masses. The origin of moduli space is therefore unstable, 
and some of the mesons or baryons must have non-vanishing vevs in the 
true vacuum. Note that our method only gives us information on the 
form of the potential close to the origin, since far from the origin operators with higher
powers of meson and baryon fields (which we have no control over) 
are unsuppressed, and therefore we can not determine the location of the true vacuum even for
small soft breakings. Nevertheless, establishing the instability of the origin has important 
consequences, since in these theories  
all points on the moduli space away from the origin 
break vector-like symmetries. If any baryonic fields obtain vevs baryon number
is broken, and if all the baryons vevs vanish, there is no point on the quantum moduli space 
where $M_{i \bar{j}} \propto \delta_{i \bar{j}}$ and so $SU(N_F)_V$ is broken. This
is to be contrasted with the non-supersymmetric theory obtained by decoupling
the scalars, where a general theorem \cite{vafawitten} shows that vector-like 
symmetries are never broken. It is easy to argue that the broken vector-like
symmetries are restored for squark masses 
larger than a (finite) critical value. Squarks of mass 
$m^2 \gg 
\Lambda^2$ can be integrated out of the theory, generating
higher dimension 
operators suppressed by $1/m^2$ in the non-supersymmetric low energy theory.
These operators can at most correct the spectrum of 
states in the 
low energy theory by $O(\Lambda^2/m^2)$. Since 
all the scalar states in the non-supersymmetric theory get masses of 
$O(\Lambda)$ (with the exception of Goldstone bosons associated with chiral 
symmetry breaking), there are no scalars which can be brought down to 
zero mass due to the $O(\Lambda^2/m^2)$ corrections and there is therefore
no candidate for the Goldstone
boson of a broken vector-like symmetry. Therefore, the vector-like symmetries
must be exactly restored above a finite critical squark mass $m^2_*
\sim \Lambda^2$,
and a phase transition must separate the nearly supersymmetric and 
non-supersymmetric theories.

For $Sp(m)$ theories with $2m + 2$ chiral multiplets, the soft mass of the 
mesons is positive and the origin of moduli space is at least a local vacuum.
At this point, the fermionic mesons are massless bound states
of a massless quark and a massive squark, the binding energy exactly 
cancelling the squark mass. 
This provides a rigorous counter-example to the 
the ``persistent mass condition" of \cite{weinberg,savas}.
 
In conclusion we remark that, while we have illustrated 
our ideas with two specific
examples, our technique for computing soft masses can clearly 
be applied in any asymptotically free supersymmetric 
theory where the theory in the deep infrared is known and is
weakly coupled, as in all $s-$confining \cite{csaba} or magnetic free theories.

NAH would like to thank S. Thomas and M. Peskin for very useful discussions. 
RR would like to thank Alberto Zaffaroni for useful conversations. RR is also
indebted to Erich Poppitz for discussions and for suggesting
the analogy with the pion mass.  Both authors acknowledge
very useful comments and criticism from Markus Luty, and useful correspondence
with H. Nishino.

\def\ijmp#1#2#3{{\it Int. Jour. Mod. Phys. }{\bf #1~}(19#2)~#3}
\def\pl#1#2#3{{\it Phys. Lett. }{\bf B#1~}(19#2)~#3}
\def\zp#1#2#3{{\it Z. Phys. }{\bf C#1~}(19#2)~#3}
\def\prl#1#2#3{{\it Phys. Rev. Lett. }{\bf #1~}(19#2)~#3}
\def\rmp#1#2#3{{\it Rev. Mod. Phys. }{\bf #1~}(19#2)~#3}
\def\prep#1#2#3{{\it Phys. Rep. }{\bf #1~}(19#2)~#3}
\def\pr#1#2#3{{\it Phys. Rev. }{\bf D#1~}(19#2)~#3}
\def\np#1#2#3{{\it Nucl. Phys. }{\bf B#1~}(19#2)~#3}
\def\mpl#1#2#3{{\it Mod. Phys. Lett. }{\bf #1~}(19#2)~#3}
\def\arnps#1#2#3{{\it Annu. Rev. Nucl. Part. Sci. }{\bf #1~}(19#2)~#3}
\def\sjnp#1#2#3{{\it Sov. J. Nucl. Phys. }{\bf #1~}(19#2)~#3}
\def\jetp#1#2#3{{\it JETP Lett. }{\bf #1~}(19#2)~#3}
\def\app#1#2#3{{\it Acta Phys. Polon. }{\bf #1~}(19#2)~#3}
\def\rnc#1#2#3{{\it Riv. Nuovo Cim. }{\bf #1~}(19#2)~#3}
\def\ap#1#2#3{{\it Ann. Phys. }{\bf #1~}(19#2)~#3}
\def\ptp#1#2#3{{\it Prog. Theor. Phys. }{\bf #1~}(19#2)~#3}

\end{document}